\documentclass{aastex631}

\usepackage{amsmath}
\usepackage{amsfonts}

\begin{document}

\title{Spin Precession Signatures as an Indicator of Microlensing in Strongly Lensed Gravitational Waves}

\author[0000-0003-3201-061X]{Xikai Shan}
\affiliation{Department of Astronomy, Tsinghua University, Beijing 100084, China}
\email{xk\_shan@mail.bnu.edu.cn}

\author[0000-0002-9965-3030]{Huan Yang}
\affiliation{Department of Astronomy, Tsinghua University, Beijing 100084, China}

\author[0000-0001-8317-2788]{Shude Mao}
\affiliation{Department of Astronomy, Westlake University, Hangzhou 310030, Zhejiang Province, China}
% \email{shude.mao@westlake.edu.cn}

\author[0000-0002-3887-7137]{Otto A. Hannuksela}
\affiliation{Department of Physics, The Chinese University of Hong Kong, Shatin, NT, Hong Kong}



\begin{abstract}
Microlensing by the stellar field in a strong-lensing galaxy can introduce wave-optics distortions into the waveforms of strongly lensed gravitational waves (SLGWs). If these signals are analyzed with waveform templates that do not include microlensing, the lensing-induced modulation may be misinterpreted as intrinsic source physics. In particular, microlensing can mimic spin precession, since both effects can produce beat-pattern-like features in the waveform. In this work, we study the degeneracy between stellar-field microlensing and spin precession, and ask to what extent microlensed SLGWs may show false evidence of precession. We analyze simulated SLGW events for two detector sensitivities, O5 and a lower-noise configuration with a power spectral density reduced by a factor of 4 (named O5 Plus), assuming binary black holes with parallel spins.
We find that microlensing can indeed produce apparent evidence for precession, and that this effect becomes more visible at higher signal-to-noise ratios. Under O5 sensitivity, 4.88\% of microlensed events lie above the one-sided Gaussian-equivalent \(3\sigma\) background threshold, corresponding to the 99.9th percentile of the unlensed-background distribution, while under O5 Plus sensitivity this fraction increases to 14.91\%. We also find that the evidence for precession is positively correlated with the strength of microlensing. This correlation is weak under O5 sensitivity, but becomes clear under O5 Plus sensitivity. In addition, Type II (saddle-point) images show a stronger correlation than Type I (minimum-point) images. These results show that evidence for precession in GW data should be interpreted with care, as it may also arise from microlensing wave effects in SLGWs.
\end{abstract}

\keywords{Gravitational wave --- Gravitational lensing --- Microlensing --- Strong lensing}

%% cross-references to sections, equations, tables, and figure.

\section{Introduction} 
\label{sec:intro}
Strongly-lensed gravitational waves (SLGWs) offer a new method for probing the Universe~\citep{Li:2018prc, Oguri:2019fix, Hannuksela:2020xor, 2022A&A...659L...5C, Liao:2022gde, PhysRevLett.130.261401, Seo_2024, 2025MNRAS.536.2212P}. 
%We should not be content with merely applying them to familiar problems, but should also explore the unique phenomena that this new probe can reveal.
Unlike strongly lensed quasars or supernovae, GWs from binary black hole coalescences can have wavelengths comparable to the Schwarzschild radius of the lens. When these waves pass through a stellar field in the lensing galaxy, additional interference and diffraction effects can arise~\citep{10.1143/PTPS.133.137, Nakamura:1997sw, Takahashi:2003ix,Diego:2019lcd, Cheung2021Stellar-massWaves, Mishra:2021xzz, Meena:2022unp, Yeung2023DetectabilityWaves, Seo:2025dto}.
By extracting the interference imprints within GWs, it may be possible to reconstruct substructure information of the lens galaxy, such as the properties of intermediate-mass black holes~\citep{Lai:2018rto} and dark matter substructures~\citep{Liu:2023ikc}.

The most critical step in the application of SLGWs is the accurate and efficient identification of such events. However, the two main methods currently available, the posterior-overlap method~\citep{Haris:2018vmn,Barsode2024FastWaves} and the joint parameter estimation (joint PE) method~\citep{Lo:2021nae,Liu:2020par,Janquart:2021nus,Janquart_2023}, are affected to varying degrees by issues related to false-positive rates~\citep{Caliskan:2023zqm,Wierda2021BeyondLensing}. 
% Although using population models to provide priors, such as time delay information, can alleviate this issue, i
Introducing a novel additional criterion would be a valuable contribution to these detection strategies. On the other hand, neglecting such wave-optics effects can lead to missed signals~\citep{Chan2024DetectabilitySearches} and biased parameter estimation for SLGWs, particularly when an event is highly magnified or passes through regions with a high density of compact lenses~\citep{Meena:2022unp,Mishra2024ExploringDetection,Shan:2023qvd}. 
However, in such cases, significant wave-optics effects can also provide additional evidence for strong lensing. Therefore, understanding wave optics is important not only for probing small-scale lenses but also for ensuring the robust detection of strongly lensed events.

As research into microlensing wave optics progresses, it has been observed that the effects of microlensing on GW waveforms bear similarities to intrinsic precession effects, with both producing beat-pattern-like modulations in the waveform (note that overlapping signals can also introduce a similar beat-pattern-like effect, which is beyond the scope of this study~\citep{Hu:2025vlp}).
Although the study by \citet{Liu:2023emk} showed that these two effects can be disentangled using template matching, this result mainly applies to cases where template construction is relatively straightforward. One example is the eikonal limit, in which the GW wavelength is much smaller than the characteristic size of the lens, so one only needs to consider the superposition of waveforms. This mainly applies to lenses above about 100 solar masses.
For a microlensing field composed of thousands to millions of stars with masses below about 10 solar masses, constructing such templates becomes extremely challenging. As a result, the ``similarity'' between the two effects cannot be easily disentangled, leading to potential degeneracies. This means that microlensing effects could be mistakenly interpreted as intrinsic precessional effects.

At the same time, although binary black holes are generally expected to precess, the observable precession can be very weak in some formation channels. As a result, clear evidence for precession remains difficult to obtain for many events with current detectors. This makes it important to ask whether some apparent evidence for precession could instead be caused by microlensing. If so, such degeneracies could bias population studies of binary black holes and lead to incorrect conclusions about their formation channels. This is because precession measurements can help distinguish among different formation scenarios, at least in a statistical sense. In particular, one may incorrectly favor black holes formed through dynamical interactions in dense clusters (the dynamical channel)~\citep{1993Natur.364..423S, 2000ApJ...528L..17P, Rodriguez_2016, Mapelli_2022}, which are more likely to produce stronger precession, over those formed through the evolution of massive binary stars (the EMBS channel)~\citep{LIGOScientific:2016aoc, Belczynski_2016, 2018MNRAS.480.2011G}, which often predicts weaker observable precession.

In addition, in the context of strong-lensing identification, these degeneracies could serve as a new clue for strongly lensed events. In other words, one may further filter candidate strong-lensing events by identifying those that show evidence for precession, thereby helping to reduce the false-positive rate. The main goal of this work is to study the possible degeneracy between microlensing and precession in GW waveforms, and to ask whether microlensing can produce observable evidence for precession. If so, evidence for precession may not only suggest a dynamical formation channel, but may also point to a microlensing candidate that deserves further investigation. We find that microlensing can indeed produce evidence for precession, and that this effect depends on both the strength of the microlensing and the detector sensitivity. In future lower-noise detectors, events with strong microlensing effects, for example highly magnified events, are also more likely to be identified as precessing binaries. Therefore, evidence for precession may serve as an additional criterion for identifying SLGWs.

% In population studies of binary black holes, mis-identification of precessional systems could lead to incorrect conclusions about the formation mechanisms of binary black hole systems, potentially favouring dynamical interaction in dense clusters (the dynamical channel)~\citep{1993Natur.364..423S, 2000ApJ...528L..17P, Rodriguez_2016, Mapelli_2022}, as compared to through the evolution of isolated binary stars (the EMBS channel)~\citep{LIGOScientific:2016aoc, Belczynski_2016, 2018MNRAS.480.2011G}. 
% On the other hand, in the context of lensing identification, these degeneracies could serve as new evidence for strongly lensed events, as every strongly-lensed image must be microlensed and more than 10\% of events will have microlensing-induced mismatches $>0.03$ in the waveform~\citep{Shan:2024min}.
% In other words, if a candidate list of strong-lensing events also show a strong correlation with spin precession, as compared to average events, the false alarm rate may be further reduced.
% %one might further filter candidate strong lensing events by identifying precession-like effects, thereby reducing the false positive rate.

The structure of the paper is as follows: in Section~\ref{sec:simulation}, we introduce the fundamental theories of strong lensing and microlensing, along with the simulation processes used for the data.
Section~\ref{sec:result} presents our results, and finally, we provide a summary and discussion in Section~\ref{sec:conclu_discus}.

\section{Basic theory and Mock Data simulation} 
\label{sec:simulation}
The GW lensing effect induced by a microlensing field embedded within a strong lensing galaxy can be described through the diffraction integral~\citep{schneider1992gravitational,10.1143/PTPS.133.137,Takahashi:2003ix}, as described in the following equation:

\begin{equation}
\label{eq:DiffInter}
F(\omega, \boldsymbol{y})=\frac{2 G \langle \mathrm{M}_L \rangle(1+z_L) \omega}{\pi c^{3} i} \int_{-\infty}^{\infty} d^{2} x \exp \left[i \omega t(\boldsymbol{x}, \boldsymbol{y})\right].
\end{equation}
Here, $F(\omega, \boldsymbol{y})$ represents the amplification factor, while $\omega$ and $\boldsymbol{y}$ denote the GW angular frequency and its position in the source plane (normalized by the Einstein radius of average microlens mass $\langle \mathrm{M}_L \rangle$), respectively. 
$z_L$ refers to the microlens redshift and $\boldsymbol{x}$ corresponds to the coordinates in the lens plane (normalized by the Einstein radius of $\langle \mathrm{M}_L \rangle$).

In Eq.~(\ref{eq:DiffInter}), the term $t(\boldsymbol{x}, \boldsymbol{y})$ represents the time delay function for the microlensing field embedded in the lens galaxy or galaxy cluster, which can be expressed as~\citep{Wambsganss1990, 1992grlebookS, 2021xuechunchen}:
\begin{equation}
\begin{split}
\label{eq:TimeDelay}
t(\boldsymbol{x},\boldsymbol{x}^{i},\boldsymbol{y}=0)&=\underbrace{\frac{k}{2}\left((1-\kappa+\gamma) x_{1}^{2}+(1-\kappa-\gamma) x_{2}^{2}\right)}_{t_\text{smooth}(\kappa,\gamma,\boldsymbol{x})}-\underbrace{\left[\frac{k}{2}\sum_{i}^{N_*} \frac{\mathrm{M}_{L,i}}{\langle \mathrm{M}_L \rangle} \ln \left(\boldsymbol{x}^{i}-\boldsymbol{x}\right)^{2} + k\phi_{-}(\boldsymbol{x})\right]}_{t_\text{micro}(\boldsymbol{x},\boldsymbol{x}^{i})} \ .
\end{split}
\end{equation}
In this equation, $k = 4 G \langle \mathrm{M}_L \rangle (1 + z_L)/c^3$, $\mathrm{M}_{L,i}$ and $\boldsymbol{x^i}$ refer to the mass and position of the $i$th microlens, respectively, and $N_*$ indicates the number of microlenses. 
The parameters $\kappa$ and $\gamma$ denote the convergence and shear of the macro lens, respectively. 
Additionally, $t_\text{smooth}(\kappa, \gamma, \boldsymbol{x})$ and $t_\text{micro}(\boldsymbol{x}, \boldsymbol{x}^i)$ represent the macro and microlensing time delays, respectively. 
For simplicity, the macro image position is set to the origin ($\boldsymbol{y} = 0$). 
A negative mass sheet, represented by $\phi_{-}(\boldsymbol{x})$, is included to ensure that the total convergence $\kappa$ remains unchanged when microlenses are added~\citep{Wambsganss1990, 2021xuechunchen, zheng2022}.

\begin{figure}
\centering
\includegraphics[width=0.5\columnwidth]{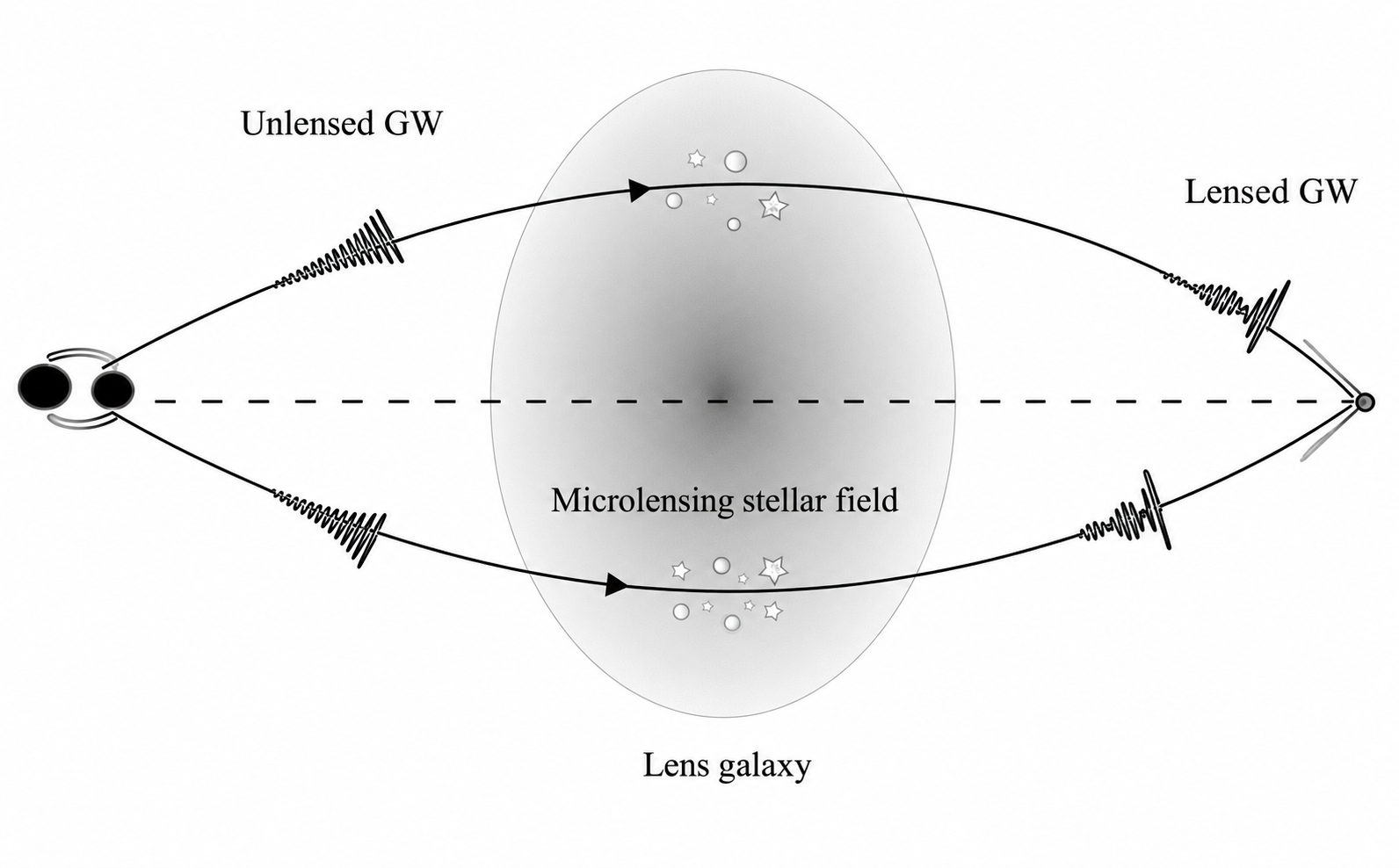}
\caption{Illustrative sketch of a SLGW influenced by a microlensing stellar field.
A GW source is depicted on the far left. 
The signal propagating towards the lens galaxy (elliptical shape) represents the unlensed waveform. Individual stars within the lens galaxy act as microlenses. 
The signal emerging after passing through the lens galaxy is the lensed waveform, observed by a GW detector on the far right.
This figure is adapted from Figure 2 in \citet{Seo:2025dto}.}
\label{fig:sketch}
\end{figure}

Figure~\ref{fig:sketch} shows an illustrative sketch of the strong lensing GW influenced by a microlensing stellar field in the lens galaxy.
One can see that the strongly-lensed  GW waveform is further  modulated by the microlensing effect.
Note that the number of stars depicted in the lens galaxy in this figure is purely illustrative and not intended to represent realistic values. 
Typically, the number of stars can range from $10^3$ to $10^6$~\citep{1986ApJ...306....2K, Shan:2022xfx}.

In this study, we adopt the same strategy as in \citet{Shan:2023ngi, Shan:2023qvd} to simulate SLGW signals. 
Specifically, we assume that the merger rate of binary black holes is proportional to the star formation rate~\citep{Haris:2018vmn}, and determine whether a GW event is a strong lensing event based on the multi-image optical depth of the singular isothermal sphere (SIS) model. 
We then calculate the microlensing stellar field density at the positions of the strong lensing images, which follow the S\'{e}rsic light profile~\citep{Vernardos_2018}.

The microlensing stellar field consists of two components: stars and remnants. 
Here, we assume that the stellar mass function follows the Chabrier initial mass function~\citep{2003PASP..115..763C}, while the remnant mass function is determined using the initial-final mass relation described in~\citet{2015MNRAS.451.4086S}. 
Finally, we use the TAAH (\textbf{T}rapezoid \textbf{A}pproximation-based \textbf{A}daptive \textbf{H}ierarchical) method proposed by \citet{Shan:2024min} to evaluate the microlensing wave effects. 
For further details, please refer to Appendix~\ref{app:Strong_micro_simul}.

Based on the previous procedure, we build two SLGW data sets: one containing events detectable at the O5 noise level~\citep{2020LRR....23....3A}, and the other containing events with the same GW waveform (i.e., the same strong-lensing and microlensing effects) but at a lower noise level, with the noise power spectrum reduced by a factor of $4$ compared to O5. Hereafter, we refer to this lower-noise configuration as O5 Plus.
Here, we assume Gaussian noise and do not include any non-Gaussian features such as glitches.
Each data set contains 125 binary black hole merger signals.
The reason we also choose a lower-noise detector is that we want to investigate the influence of the signal-to-noise ratio (SNR) on the detectability of microlensing-induced degeneracy and to see whether the precession evidence induced by microlensing could be a useful criterion for SLGWs in the future.
Here, we use the IMRPhenomXP~\citep{Pratten_2021} waveform model with aligned-spin parameters to generate GW signals.
We do not include intrinsic precession in the mock data.
This is because our main goal is to study the degeneracy between microlensing and precession.
To do this clearly, it is better to start with signals that have no true precession, so that any recovered evidence for precession can be directly attributed to microlensing.
If the signal already contains intrinsic precession, then precession evidence would be easier to obtain, because the precessing part of the signal cannot be fully captured by aligned-spin templates.
In that case, it would be harder to isolate the specific effect caused by microlensing.

\section{Result} 
\label{sec:result}
\subsection{Microlensing induced precession evidence}
\label{subsec:evidence}
In this section, we characterize the degree of degeneracy between microlensing and precession by quantifying the precession Bayes factor induced by microlensing effects in SLGW signals. Specifically, in the parameter-estimation procedure, we recover the GW parameters using the Dynesty~\citep{2019S&C....29..891H} sampler under two different assumptions, while using the same waveform template, IMRPhenomXP. In one case, the template includes precession effects, and in the other, it assumes parallel spins. We then compute the Bayes factor for precession by comparing the detection evidences under these two assumptions.

To quantify the strength of the precession evidence induced by microlensing, we also simulate a background sample. This sample consists of 125 standard GW events under O5 sensitivity, together with 125 additional standard GW events with the noise power spectrum reduced by a factor of 4 relative to O5. Here, ``standard'' refers to events drawn from the same BBH population model but without strong lensing or spin precession. We then define two background-calibrated thresholds based on the ranking statistic. Specifically, for a reference significance level $s$, we define the threshold as the $\Phi(s)$ quantile of the unlensed-background distribution, where $\Phi$ is the cumulative distribution function of the standard normal distribution. In this work, we use $s=1$ and $s=3$, corresponding to one-sided Gaussian-equivalent reference levels of $1\sigma$ and $3\sigma$, or equivalently to the 84.1st and 99.9th percentiles of the background distribution, respectively.

\begin{figure} 
\centering 
\includegraphics[width=\columnwidth]{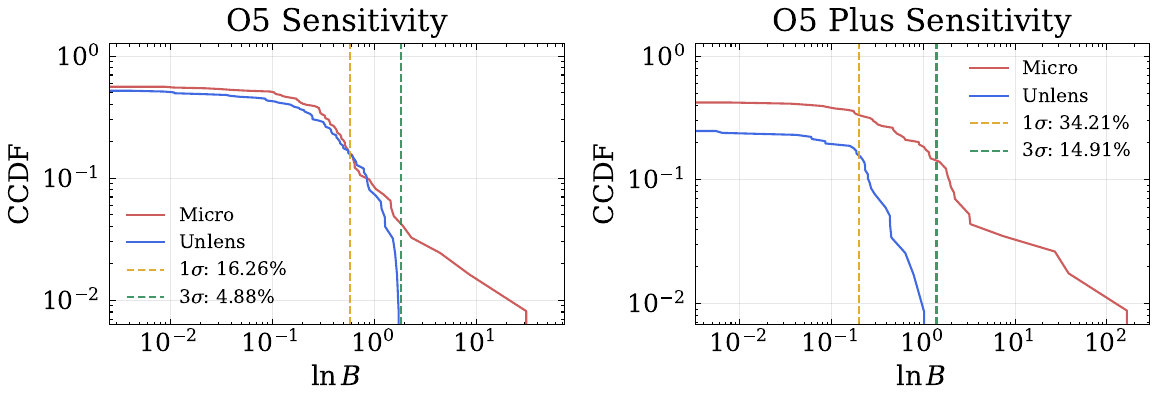} 
\caption{Complementary cumulative distribution functions (CCDFs) of the logarithmic Bayes factor for precession for microlensed events (red curves) and normal unlensed events (blue curves). The left and right panels correspond to O5 and O5 Plus sensitivities, respectively. The dashed vertical lines mark the one-sided Gaussian-equivalent background thresholds, defined as the $\Phi(1)$ and $\Phi(3)$ quantiles of the unlensed-background distribution, corresponding to one-sided Gaussian significance levels of $1\sigma$ and $3\sigma$, respectively.}
\label{fig:bayes_CCDF} 
\end{figure}

Figure~\ref{fig:bayes_CCDF} shows the complementary cumulative distribution function (CCDF), defined as $P(\ln B>x)$, of the logarithmic Bayes factor for precession for both microlensed events (red curve) and normal unlensed events (blue curve). The left panel shows the results for O5 sensitivity, and the right panel shows the results for O5 Plus sensitivity. We use two dashed vertical lines to mark the $1\sigma$- and $3\sigma$-equivalent background thresholds. We find that, in the lower-SNR scenario (O5), 16.25\% and 4.88\% of events lie above the $1\sigma$- and $3\sigma$-equivalent background thresholds, respectively. In the higher-SNR scenario (O5 Plus), these fractions increase to 34.21\% and 14.91\%, respectively.

The first main reason for the higher fraction at higher SNR is that, in the linear-signal approximation, the Bayes factor scales as~\citep{Cornish_2011}
\begin{equation}
\label{eq:lnB}
\ln B = \rho^{2}(1-\mathrm{FF}^2)/2+\mathrm{const},
\end{equation}
where $\rho$ is the SNR and $\mathrm{FF}$ is the fitting factor, defined as
\begin{equation}
\label{eq:F_F}
\mathrm{FF} = \max_{\theta} \frac{\langle {h}_1 \mid {h}_2\rangle}{\sqrt{\langle {h}_1 \mid {h}_1\rangle\langle {h}_2 \mid {h}_2\rangle}}\,, 
\end{equation}
where $h_1$ is the unlensed waveform and $h_2$ is the microlensed waveform. The ${\theta}$ denotes all GW parameters. The operator $\langle \cdot \mid \cdot \rangle$ represents the noise-weighted inner product, defined as
\begin{equation}
\langle {h}_{1} \mid {h}_{2}\rangle=4 \operatorname{Re} \int_{f_{\text {low}}}^{f_{\text {high}}} \mathrm{d} f \frac{{h}_{1}^*(f) {h}_{2}(f)}{S_{\mathrm{n}}(f)}\,,
\end{equation}
where $S_\mathrm{n}(f)$ is the single-sided power spectral density of the detector noise, and ``$^*$'' denotes complex conjugation. Therefore, for a fixed $\mathrm{FF}$, the Bayes factor increases with $\rho^{2}$.

\begin{figure}
\centering
\includegraphics[width=0.5\columnwidth]{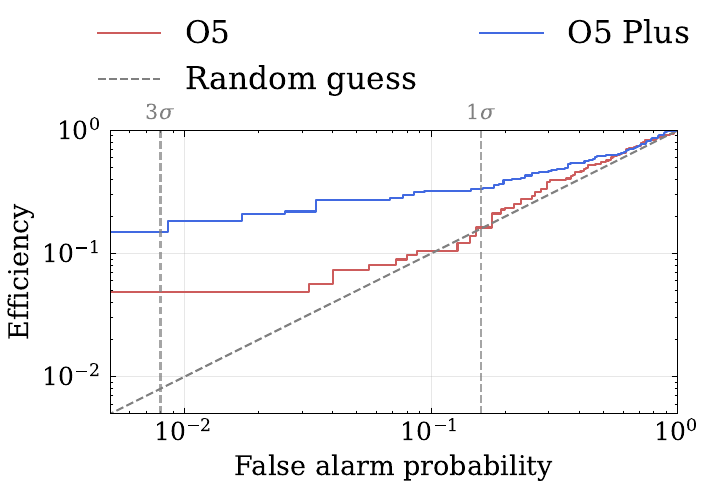}
\caption{Detection efficiency as a function of false alarm probability. The red and blue curves show the results for O5 and O5 Plus sensitivities, respectively. The grey dashed diagonal line shows the performance of random guessing. The grey dashed vertical lines mark the corresponding one-sided Gaussian-equivalent $1\sigma$ and $3\sigma$ background thresholds.}
\label{fig:ROC}
\end{figure}

The second main reason is that false-positive support for precession from the unlensed background is suppressed at higher SNR.
For nested models, the non-precessing hypothesis is a special case of the precessing hypothesis obtained at the non-precessing limit of the additional precession parameters. 
In this case, the Bayes factor in favour of the simpler non-precessing model over the precessing model can be written through the Savage--Dickey density ratio as \citep{Wagenmakers2010}
\begin{equation}
B_{\mathrm{non\mbox{-}prec}/\mathrm{prec}}
=
\frac{p(\phi=0 \,|\, d,\mathrm{prec})}{p(\phi=0 \,|\, \mathrm{prec})}
\end{equation}
where $\phi$ denotes the additional precession parameters, and $\phi=0$ corresponds to the non-precessing limit. 
Here, the numerator is the posterior density at the non-precessing point under the precessing model, while the denominator is the corresponding prior density. 
Therefore, the Bayes factor in favour of precession is
\begin{equation}
B_{\mathrm{prec}/\mathrm{non\mbox{-}prec}}
=
B_{\mathrm{non\mbox{-}prec}/\mathrm{prec}}^{-1}.
\end{equation}
For a truly unlensed and non-precessing event, the data do not contain a genuine precession signal, so any preference for the precessing model can only arise from fitting random noise fluctuations. 
In the high-SNR limit, the Fisher matrix for the additional precession parameters scales as $\Gamma_{\phi\phi}\propto \rho^{2}$, so the corresponding covariance scales as $\Gamma_{\phi\phi}^{-1}\propto \rho^{-2}$. 
Therefore, the characteristic posterior width of each independently measurable additional parameter scales as $\Delta\theta\propto \rho^{-1}$. 
Hence, for $k_{\mathrm{eff}}$ effective additional precession degrees of freedom, the posterior volume in the precession sector shrinks as
\begin{equation}
V_{\mathrm{post}}(\phi) \propto \rho^{-k_{\mathrm{eff}}},
\end{equation}
which implies
\begin{equation}
p(\phi=0 \mid d,\mathrm{prec}) \propto \rho^{k_{\mathrm{eff}}},
\end{equation}
while the prior density $p(\phi=0 \mid \mathrm{prec})$ is independent of $\rho$. 
Therefore,
\begin{equation}
B_{\mathrm{non\mbox{-}prec}/\mathrm{prec}}^{\mathrm{bg}} \propto \rho^{k_{\mathrm{eff}}},
\qquad
B_{\mathrm{prec}/\mathrm{non\mbox{-}prec}}^{\mathrm{bg}} \propto \rho^{-k_{\mathrm{eff}}},
\end{equation}
and one can expect the scaling
\begin{equation}
\label{eq:false_alarm}
\ln B_{\mathrm{bg}}
\equiv
\ln B_{\mathrm{prec}/\mathrm{non\mbox{-}prec}}^{\mathrm{bg}}
\simeq
-k_{\mathrm{eff}} \ln \rho + \mathrm{const}.
\end{equation}
Therefore, as the SNR increases, the posterior volume of the additional precession parameters shrinks and the Occam penalty becomes stronger, which suppresses the false-positive precession evidence from unlensed background events. 
Combining the above two reasons, the overlap between the microlensed and unlensed populations is reduced, leading to a higher detectability of microlensing-induced precession evidence at higher SNR.
This result also highlights the importance of waveform-model systematics for future, more sensitive GW detectors~\citep{2020PhRvR...2b3151P}.

To show more clearly how well microlensing-induced precession evidence can be separated from the unlensed background, Figure~\ref{fig:ROC} also shows the detection efficiency as a function of the false alarm probability. Here, the detection efficiency is defined as the fraction of microlensed events above a given threshold in $\ln B$, while the false alarm probability is defined as the fraction of unlensed background events above the same threshold. One can see that, for O5 Plus sensitivity, the detection-efficiency curve lies clearly above the dashed curve, which represents random guessing, i.e., the case in which microlensed and unlensed events cannot be distinguished and events are classified no better than by chance.
However, for the lower O5 sensitivity, microlensing-induced precession events can be distinguished from the background only in the long tail.

\subsection{Correlation between microlensing and precession}
\label{subsec:corre}

\begin{figure}
\centering
\includegraphics[width=\columnwidth]{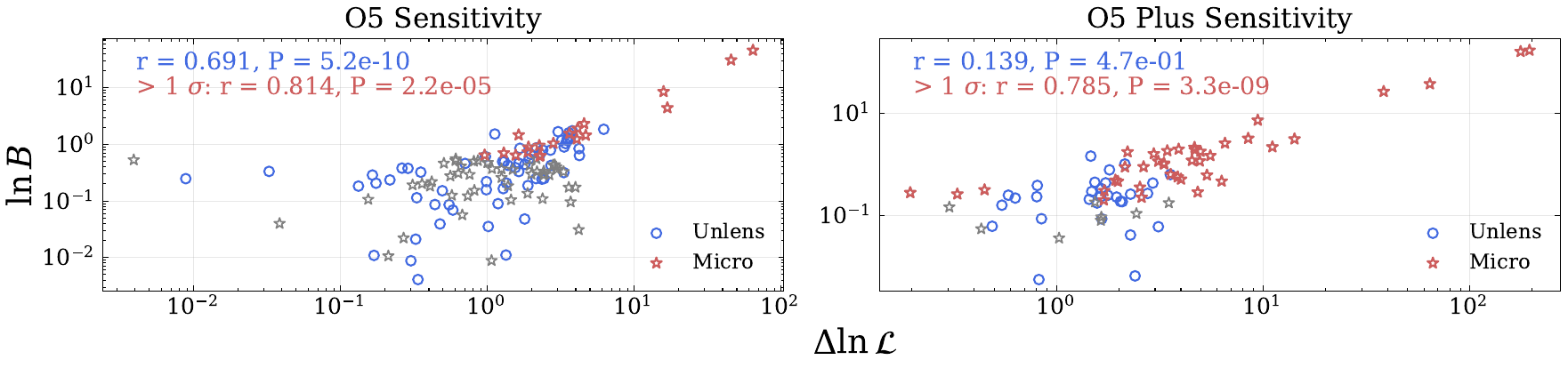}
\caption{Correlation (Spearman correlation coefficient) between the improvement in the maximum likelihood, measured between the precession and non-precession waveform inferences, and the logarithmic Bayes factor for precession. The left and right panels show the results under O5 and O5 Plus sensitivities, respectively. The blue circles represent the background events (unlensed), and the stars represent the microlensing events. Grey indicates events below the corresponding one-sided Gaussian-equivalent $1\sigma$ background threshold, while red indicates events above it. The legend gives the Spearman correlation coefficient and the $p$-value for both the background events and the microlensing events. For the microlensing events, only events above the corresponding one-sided Gaussian-equivalent $1\sigma$ background thresholds are included in the correlation calculation.}
\label{fig:dlogL_dlogB}
\end{figure}

The preceding results indicate a degeneracy between microlensing and precession. 
To further explain why the precession waveform gives higher evidence, we examine the correlation between the improvement in the maximum likelihood, $\Delta \ln \mathcal{L}$, and the logarithmic Bayes factor for precession, $\ln B$, as shown in Figure~\ref{fig:dlogL_dlogB}. Under both O5 and O5 Plus sensitivities, the microlensing events show a strong positive correlation between $\Delta \ln \mathcal{L}$ and $\ln B$. This suggests that the higher evidence for precession in some microlensing events mainly comes from the better fit of the precession waveform to the data. In this sense, the result provides a waveform-fitting explanation for the degeneracy between microlensing and precession. We also note that, under O5 sensitivity, the background unlensed events still show a moderate correlation between $\Delta \ln \mathcal{L}$ and $\ln B$. This suggests that, at lower SNR, noise can more easily produce false evidence for precession. We stress that this does not imply any physical degeneracy between precession and Gaussian noise. Rather, it reflects a statistical effect in the inference, in which the additional freedom of the precessing waveform can partially absorb random noise fluctuations and lead to an artificially enhanced preference for precession in some realizations. However, under O5 Plus sensitivity, this effect is no longer significant.

Additionally, one may ask whether larger values of the precession evidence, $\ln B$, correspond to a stronger microlensing signature. If so, $\ln B$ could be used to infer the strength of microlensing. We therefore investigate the correlation between the microlensing strength and the precession evidence, $\ln B$. Here, we quantify the microlensing strength using the mismatch induced by microlensing in the GW waveform. The mismatch is defined as follows:

\begin{equation}
\label{eq:mismatch}
\mathrm{mismatch} = 1 - \max_{\phi_0, t_0} \frac{\langle {h}_1 \mid {h}_2\rangle}{\sqrt{\langle {h}_1 \mid {h}_1\rangle\langle {h}_2 \mid {h}_2\rangle}}\,, 
\end{equation}
where $h_1$ and $h_2$ represent the waveforms of signals 1 and 2, respectively. 
$\phi_0$ and $t_0$ are the initial phase and start time of signal 1. 
This equation accounts for the time delays caused by both macrolensing and microlensing. 
In this analysis, signal 1 represents the unlensed waveform, while signal 2 represents the microlensed waveform.
Note that this match calculation is different from the fitting factor calculation (Eq.~\ref{eq:F_F}), since the former maximizes only over the initial phase and start time, while the latter maximizes over all GW parameters.

\begin{figure}
\centering
\includegraphics[width=\columnwidth]{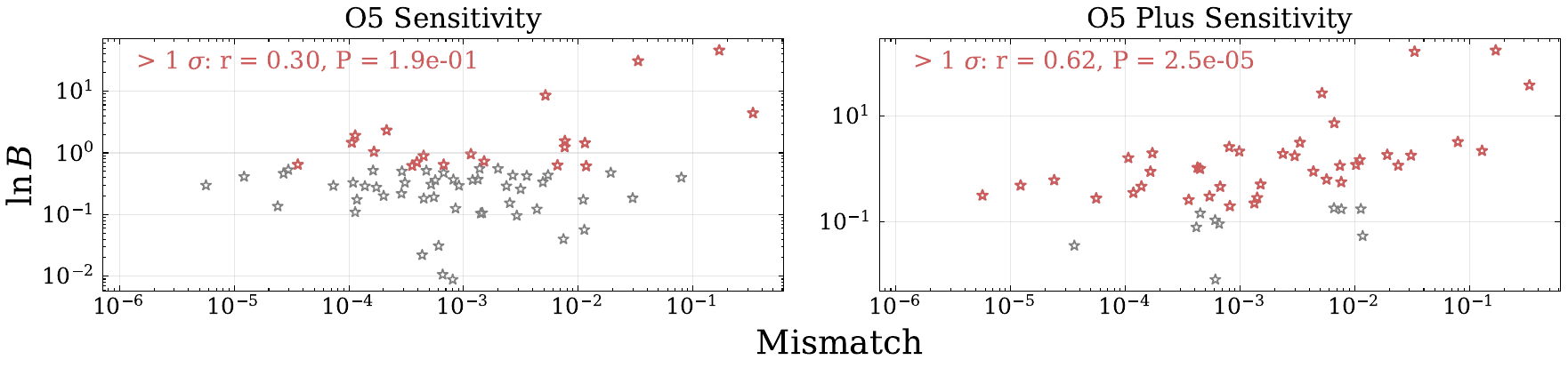}
\caption{Correlation (Spearman correlation coefficient) between the Bayes factor and the microlensing mismatch for events above the corresponding one-sided Gaussian-equivalent $1\sigma$ background thresholds, shown as red stars. The left panel shows the O5 sensitivity, while the right panel shows the O5 Plus sensitivity. The legend indicates the Spearman correlation coefficient $r$ and the $p$-value.}
\label{fig:lnb_vs_mismatch}
\end{figure}

To quantify the correlation between microlensing and precession, we also use Spearman’s correlation coefficient. Specifically, we aim to determine whether stronger microlensing effects generally lead to higher evidence for precession, or conversely, whether higher evidence for precession corresponds to a stronger microlensing effect. Figure~\ref{fig:lnb_vs_mismatch} shows the distribution of $\ln B$ for SLGW events as a function of microlensing mismatch. 
The left panel shows the events under O5 sensitivity, while the right panel shows those under O5 Plus sensitivity. Here, we use only the events above the corresponding one-sided Gaussian-equivalent $1\sigma$ background thresholds, shown as red stars, to calculate the correlation coefficient. One can find that, under O5 sensitivity, there is only a weak correlation between the mismatch and $\ln B$. This means that such events are strongly affected by noise, and a stronger microlensing effect does not necessarily lead to a higher $\ln B$. However, in the right panel, under the higher-SNR configuration, the correlation coefficient reaches 0.62 and the $p$-value is as low as the order of $10^{-5}$, showing a high level of statistical significance. This means that a stronger microlensing effect often corresponds to higher evidence for precession. Therefore, if one detects an event with very high precession evidence, then, besides the possibility that it is a true precession event, it may also be a lensing event with strong microlensing.

\begin{figure}
\centering
\includegraphics[width=\columnwidth]{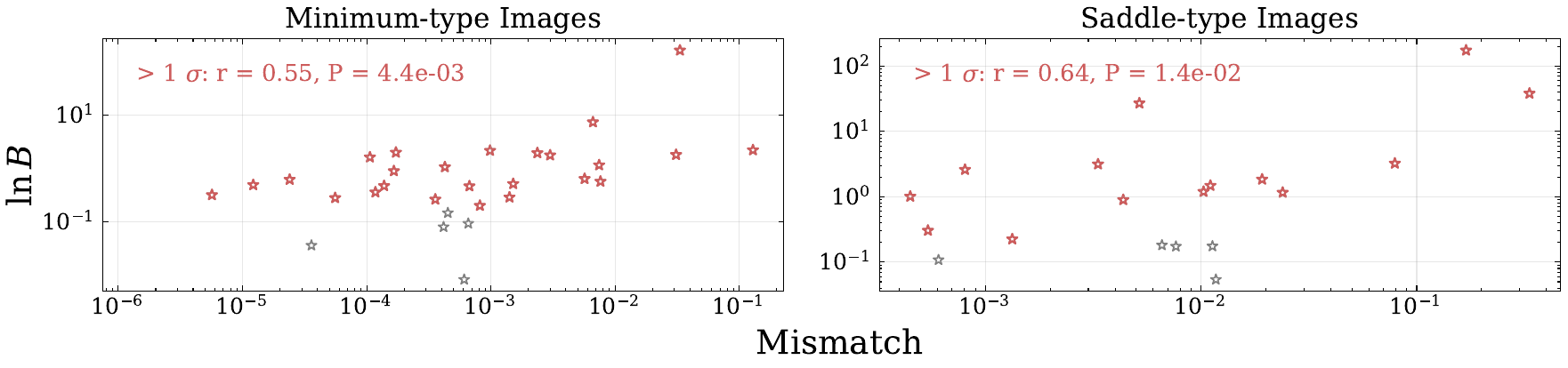}
\caption{Correlation (Spearman correlation coefficient) between the Bayes factor and the microlensing mismatch for different types of strong-lensing images. The left panel shows the minimum point, while the right panel shows the saddle point. Red stars indicate events above the corresponding one-sided Gaussian-equivalent $1\sigma$ background thresholds, while grey points indicate events below them. The legend indicates the Spearman correlation coefficient $r$ and the $p$-value.}
\label{fig:lnb_vs_mismatch_min_sad}
\end{figure}

Finally, we investigate the correlation between microlensing strength and precession evidence for two types of SLGW images: Type I, originating from the minimum point of the time-delay surface, and Type II, originating from the saddle point of the time-delay surface. We then ask which type of event can more easily mimic a precession signature. In this analysis, we only consider events under O5 Plus sensitivity, as shown in the right panel of Figure~\ref{fig:lnb_vs_mismatch}. The results are shown in Figure~\ref{fig:lnb_vs_mismatch_min_sad}.

The left panel of the figure corresponds to Type I images, while the right panel corresponds to Type II images. Here, we also restrict the analysis to events above the corresponding one-sided Gaussian-equivalent $1\sigma$ background thresholds in order to reduce the impact of false positives caused by noise.

We find that the correlation is slightly stronger for Type II images than for Type I images. The correlation coefficients for Type I and Type II images are 0.55 and 0.64, respectively, and the $p$-values are of the order of $10^{-3}$ and $10^{-2}$, showing statistical significance at this level. Therefore, we conclude that, for each strong-lensing image type, there is a moderate positive correlation between the strength of the microlensing effect and the evidence for precession. Thus, one can expect stronger evidence for precession when the microlensing effect is stronger.

These results also suggest that microlensing effects in Type II images can more easily mimic precession effects. For a better understanding, we can return to Eq.~(\ref{eq:lnB}). This scaling relation indicates that, from a statistical point of view, Type II images will have a smaller fitting factor than Type I images at a comparable SNR. This also suggests that orbital precession is more effective at capturing the microlensing-induced fluctuations in Type II events than in Type I events. This is likely because the microlensing effects in Type II images produce a stronger modulation of the phase. The phase modulation between Type I and Type II images can be compared in the rightmost panel of Figure~\ref{fig:Ft_Ff_28_}, as shown there. Therefore, for one SLGW pair, one may expect to observe one event with a precession effect and another without a precession effect. Consequently, when identifying SLGW events using the parameter-overlap method or the joint parameter-estimation method, we need to treat these precession parameters with care.

\section{Conclusion and discussion}
\label{sec:conclu_discus}

Wave-optical effects in gravitational lensing can produce frequency-dependent fluctuations in GW waveforms, often described as a ``beat pattern''~\citep{Diego:2019lcd, PhysRevD.101.064011}. Similar features can also arise from intrinsic source physics, especially orbital precession in BBH systems. This similarity makes it important to understand under what conditions microlensing can mimic precession, and how this may affect the interpretation of GW signals. In previous work, \citet{Liu:2023emk} showed that these two effects can in principle be separated with a dedicated template that includes both lensing under geometric approximation and precession. 
However, that strategy is much harder to apply in a stellar microlensing field, where the number of stars and remnants can be as large as $10^6$~\citep{1986ApJ...306....2K}. In such a situation, constructing a complete template becomes very difficult.
Recent work has suggested that simulation-based inference (SBI) may provide a possible way forward for this problem, because they can learn the statistical mapping between waveform features and the underlying microlensing field without requiring a fully complete template~\citep{Su:2025xry}. 
However, such methods are still at an early stage, and it remains valuable to study how microlensing can be confused with intrinsic binary parameters in standard MCMC-type inference settings.

In this work, we studied how often microlensing wave effects in a strongly lensed galaxy can produce precession-like signatures, and how the inferred precession evidence is related to the strength of microlensing. To do this, we analyzed two SLGW datasets with different detector sensitivities, corresponding to O5 and a lower-noise O5 Plus configuration. 
Our results show that microlensing can indeed generate apparent evidence for precession, and that this effect becomes more visible at higher SNR. Under O5 sensitivity, 16.25\% and 4.88\% of microlensed events lie above the one-sided Gaussian-equivalent \(1\sigma\) and \(3\sigma\) background thresholds, respectively. Under O5 Plus sensitivity, these fractions increase to 34.21\% and 14.91\%, respectively. The detection efficiency in the higher-sensitivity case is also clearly above the level expected from random guessing, as shown in Figure~\ref{fig:ROC}. This means that, in a quieter detector environment, the precession-like signature produced by microlensing is easier to identify.

The physical reason is twofold. First, for a fixed fitting factor, the Bayes factor grows with SNR, so waveform differences caused by microlensing are more easily translated into measurable evidence at higher SNR. Second, the false-positive support for precession from unlensed background events is reduced as the SNR increases. Together, these two effects decrease the overlap between microlensed and unlensed populations and improve the detectability of microlensing-induced precession-like signals. This point is important for future detectors with even higher sensitivity, where waveform-model systematics, such as microlensing modulation, will become more important and can bias the inference and produce false evidence for intrinsic parameters. However, such precession-like signatures may also provide useful supplementary information for identifying SLGW events.

We also examined whether stronger microlensing tends to produce larger precession evidence. The answer depends on the detector sensitivity. Under O5 sensitivity, the correlation between the microlensing mismatch and $\ln B$ is weak, suggesting that noise still plays a large role and can hide the underlying trend, as also shown by the unlensed events in Figure~\ref{fig:dlogL_dlogB}. Under O5 Plus sensitivity, however, the correlation becomes much clearer, with a Spearman coefficient of 0.62 and a $p$-value of order $10^{-5}$. This indicates that, in a higher-SNR regime, stronger microlensing usually leads to stronger evidence for precession. Therefore, a GW event with large precession evidence does not always have to be a truly precessing system; it may also be a lensed event with a strong microlensing wave effect.

We further compared two types of SLGW images, Type I and Type II. We found that Type II images show a slightly stronger correlation between microlensing strength and precession evidence than Type I images. This suggests that Type II images are more likely to mimic precession. A likely reason is that microlensing produces stronger phase modulation in Type II images, making them easier to fit with a precessing waveform. As a result, within one SLGW pair, it is possible that one image shows apparent precession while the other does not. This possibility should be treated carefully when SLGW candidates are identified using parameter-overlap methods or joint parameter-estimation methods.

Overall, our results show that precession evidence in GW data should be interpreted with care. Strong evidence for precession may point to a genuinely precessing BBH and may therefore support a dynamical formation channel, but it may also be produced by microlensing wave effects in an SLGW event. Since clear precession evidence appears to be rare in the present GW catalog, such events deserve special attention. In some cases, precession may serve as a useful supplementary signature of SLGWs, especially for highly magnified events with large microlensing mismatches. This also means that GW events in the mass gap that show strong precession-like features may remain consistent with the possibility of being high-magnification SLGW events.

%% Also note that the akcnowlodgment environment does not support long amounts of text. If you have a lot of people and institutions to acknowledge, do not use this command. Instead, create a new \section{Acknowledgments}.
\begin{acknowledgments}
All code and results used in this work are publicly available at \url{https://github.com/xkshan97/Microlensing_Precession}.
This work is partly supported by the National Science Foundation of China (Grant No. 12133005).
X.S. acknowledges support from Shuimu Tsinghua Scholar Program (No. 2024SM199) and the China Postdoctoral Science Foundation (Certificate Number: 2025M773189).
O.A.H. acknowledges suport by grants from the Research Grants Council of Hong Kong (Project No. CUHK 14304622, 14307923, and 14307724), the start-up grant from the Chinese University of Hong Kong, and the Direct Grant for Research from the Research Committee of The Chinese University of Hong Kong. 
This material is based upon work supported by NSF's LIGO Laboratory which is a major facility fully funded by the National Science Foundation.
\end{acknowledgments}


\vspace{5mm}




\appendix

\section{Simulation procedures}
\label{app:Strong_micro_simul}
Complementing Section~\ref{sec:simulation}, this appendix details the simulation procedures used to generate a mock dataset. 
Using the Monte Carlo method and following the procedure described in~\citet{Haris:2018vmn, Xu:2021bfn, Shan:2023qvd}, we synthesized a population of GW events.
The key simulation steps were:

\textbf{Source Parameterization}: 
We use the IMRPhenomXP waveform approximation to generate a population of simulated BBH events with parameters defined as follows:
\begin{itemize}
    \item The source redshift ($z_s$) is sampled from a star formation rate (SFR) based merger model with a 50 Myr delay (see Appendix B of \citet{Xu:2021bfn}). The component masses ($m_1, m_2$) are drawn from a power-law plus peak distribution~\citep{LIGOScientific:2018jsj}.
    \item The dimensionless spin magnitudes are drawn from $a_1, a_2 \sim \mathrm{U}(0, 0.99)$.
    \item The sky location and orientation angles are sampled from their isotropic distributions:
    \begin{itemize}
        \item Inclination: $p(\iota)\propto \sin(\iota)$ for $\iota \in [0, \pi]$.
        \item Declination: $p(\delta)\propto \cos(\delta)$ for $\delta \in [-\pi/2, \pi/2]$.
        \item Right Ascension: $\alpha \sim \mathrm{U}(0, 2\pi)$.
        \item Polarization: $\psi \sim \mathrm{U}(0, \pi)$.
    \end{itemize}
    \item The merger time ($t_c$) is drawn uniformly over a 1-year period.
\end{itemize}

\textbf{Strong Lensing Selection}: The likelihood of an sBBH at redshift $z_s$ undergoing strong lensing (multiple imaging) was evaluated using the SIS optical depth, $\tau(z_s)$~\citep{Haris:2018vmn}. A stochastic selection was performed: if $\tau(z_s)$ exceeded a random variate drawn from $U(0,1)$, the event was flagged as an SLGW; otherwise, it was discarded.

\textbf{Lens Modeling for SLGWs}: Lensing effects for candidates were computed using an SIE model~\citep{1994A&A...284..285K}, with lens properties ($\sigma_v, q$) drawn from SDSS galaxy survey statistics~\citep{Choi_2007, Wierda:2021upe}.

\textbf{Observational Filtering}: Detectability was assessed using a three-detector network (LIGO Livingston/Hanford, Virgo) and an SNR threshold of 12. The simulation proceeded until 125 detectable SLGW events were obtained.

\textbf{Microlensing Environment Simulation}: The final stage involved modeling the microlensing environment pertinent to each detected SLGW. This included defining a stellar mass function based on the Chabrier IMF~\citep{2003PASP..115..763C} for stars in the mass range $[0.1, 1.5] M_\odot$~\citep{Diego:2021mhf}, and assuming an elliptical S\'{e}rsic profile~\citep{Vernardos_2018} for their density. A population of remnant objects was also incorporated, using an initial-final mass relation~\citep{2015MNRAS.451.4086S} for their masses and assuming their mass density constitutes $10\%$ of the stellar mass density.

\begin{figure}
\centering
\includegraphics[width=\columnwidth]{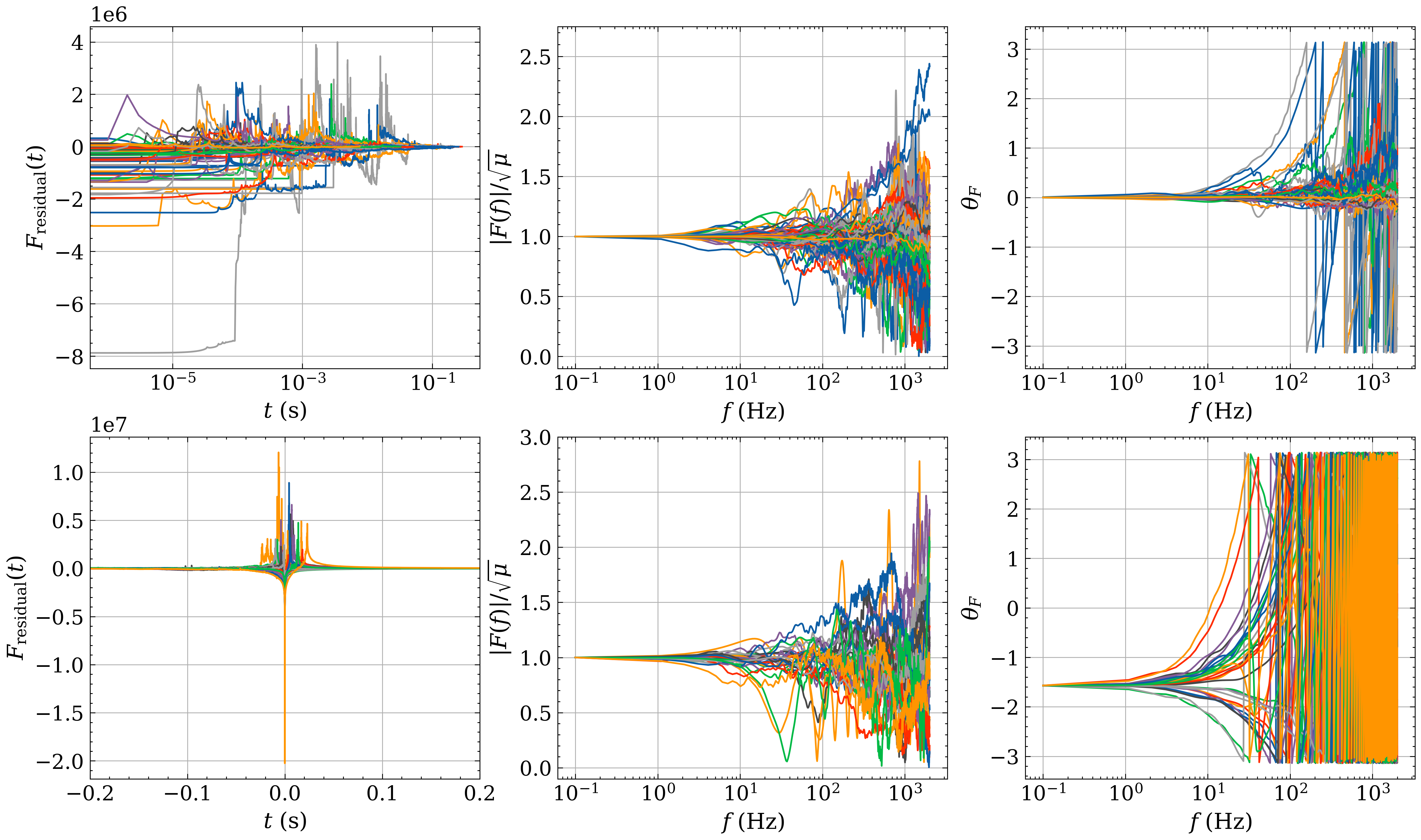}
\caption{Microlensing diffraction results for 125 SLGW events observed by a LIGO-Virgo detector network. 
Each curve corresponds to a different event. The first and second rows depict the results for Type I and Type II SLGWs, respectively. The first, second, and third columns display the residual time-domain amplification factor ($F(t) - F_{\text{smooth}}(t)$, where $F_{\text{smooth}}(t)$ excludes microlensing effects), the normalized frequency-domain amplification factor, and the complex phase, respectively.}
\label{fig:Ft_Ff_28_}
\end{figure}

Figure~\ref{fig:Ft_Ff_28_} presents the microlensing diffraction results for these simulated SLGW events.
Different curves represent different events.
The first and second rows illustrate results for Type I and Type II SLGWs, respectively. The first column shows the residual time-domain amplification factor ($F(t) - F_\mathrm{smooth}(t)$, where $F_\mathrm{smooth}(t)$ excludes microlensing effects). The second and third columns display the normalized frequency-domain amplification factor and the complex phase. The gradual convergence of each curve's tail towards zero in the first column indicates the convergence of the diffraction integral~\citep{Shan:2022xfx}.


\bibliography{sample631,references}{}

@article{Wierda2021BeyondLensing,
    title = {{Beyond the Detector Horizon: Forecasting Gravitational-Wave Strong Lensing}},
    year = {2021},
    journal = {The Astrophysical Journal},
    author = {Wierda, A. Renske A. C. and Wempe, Ewoud and Hannuksela, Otto A. and Koopmans, Léon V. E. and Broeck, Chris Van Den},
    number = {2},
    month = {11},
    pages = {154},
    volume = {921},
    publisher = {IOP Publishing},
    url = {https://iopscience.iop.org/article/10.3847/1538-4357/ac1bb4 https://iopscience.iop.org/article/10.3847/1538-4357/ac1bb4/meta},
    doi = {10.3847/1538-4357/AC1BB4},
    issn = {0004-637X},
    arxivId = {2106.06303}
}

@article{Chan2024DetectabilitySearches,
    title = {{Detectability of Lensed Gravitational Waves in Matched-Filtering Searches}},
    year = {2024},
    author = {Chan, Juno C. L. and Seo, Eungwang and Li, Alvin K. Y. and Fong, Heather and Ezquiaga, Jose M.},
    month = {11},
    url = {https://arxiv.org/abs/2411.13058v1},
    arxivId = {2411.13058}
}

@article{Yeung2023DetectabilityWaves,
    title = {{Detectability of microlensed gravitational waves}},
    year = {2023},
    journal = {Monthly Notices of the Royal Astronomical Society},
    author = {Yeung, Simon M.C. and Cheung, Mark H.Y. and Seo, Eungwang and Gais, Joseph A.J. and Hannuksela, Otto A. and Li, Tjonnie G.F.},
    number = {2},
    month = {9},
    pages = {2230--2240},
    volume = {526},
    publisher = {Oxford Academic},
    url = {https://dx.doi.org/10.1093/mnras/stad2772},
    doi = {10.1093/MNRAS/STAD2772},
    issn = {0035-8711}
}

@article{Mishra2024ExploringDetection,
    title = {{Exploring the impact of microlensing on gravitational wave signals: Biases, population characteristics, and prospects for detection}},
    year = {2024},
    journal = {Monthly Notices of the Royal Astronomical Society},
    author = {Mishra, Anuj and Meena, Ashish Kumar and More, Anupreeta and Bose, Sukanta},
    number = {1},
    month = {5},
    pages = {764--787},
    volume = {531},
    publisher = {Oxford Academic},
    url = {https://dx.doi.org/10.1093/mnras/stae836},
    doi = {10.1093/MNRAS/STAE836},
    issn = {0035-8711}
}

@article{Barsode2024FastWaves,
    title = {{Fast and efficient Bayesian method to search for strongly lensed gravitational waves}},
    year = {2024},
    author = {Barsode, Ankur and Goyal, Srashti and Ajith, Parameswaran},
    month = {12},
    url = {https://arxiv.org/abs/2412.01278v1},
    arxivId = {2412.01278}
}

@article{Cheung2021Stellar-massWaves,
    title = {{Stellar-mass microlensing of gravitational waves}},
    year = {2021},
    journal = {Monthly Notices of the Royal Astronomical Society},
    author = {Cheung, Mark H.Y. and Gais, Joseph and Hannuksela, Otto A. and Li, Tjonnie G.F.},
    number = {3},
    month = {4},
    pages = {3326--3336},
    volume = {503},
    publisher = {Oxford Academic},
    url = {https://dx.doi.org/10.1093/mnras/stab579},
    doi = {10.1093/MNRAS/STAB579},
    issn = {0035-8711},
    arxivId = {2012.07800}
}

@article{Wagenmakers2010,
title = {Bayesian hypothesis testing for psychologists: A tutorial on the Savage–Dickey method},
journal = {Cognitive Psychology},
volume = {60},
number = {3},
pages = {158-189},
year = {2010},
issn = {0010-0285},
doi = {https://doi.org/10.1016/j.cogpsych.2009.12.001},
url = {https://www.sciencedirect.com/science/article/pii/S0010028509000826},
author = {Eric-Jan Wagenmakers and Tom Lodewyckx and Himanshu Kuriyal and Raoul Grasman}
}

@article{Su:2025xry,
    author = "Su, Zhaoqi and Shan, Xikai and Lyu, Zhenwei and Zhang, Junyao and Liu, Yebin and Mao, Shude and Yang, Huan",
    title = "{From Stars to Waves: Non-deterministic Inference of Microlensed Gravitational Waves}",
    eprint = "2510.17125",
    archivePrefix = "arXiv",
    primaryClass = "gr-qc",
    month = "10",
    year = "2025"
}

@ARTICLE{2020PhRvR...2b3151P,
       author = {{P{\"u}rrer}, Michael and {Haster}, Carl-Johan},
        title = "{Gravitational waveform accuracy requirements for future ground-based detectors}",
      journal = {Physical Review Research},
         year = 2020,
        month = may,
       volume = {2},
       number = {2},
          eid = {023151},
        pages = {023151},
          doi = {10.1103/PhysRevResearch.2.023151},
archivePrefix = {arXiv},
       eprint = {1912.10055},
 primaryClass = {gr-qc},
       adsurl = {https://ui.adsabs.harvard.edu/abs/2020PhRvR...2b3151P}
}

@article{Cornish_2011,
   title={Gravitational wave tests of general relativity with the parameterized post-Einsteinian framework},
   volume={84},
   ISSN={1550-2368},
   url={http://dx.doi.org/10.1103/PhysRevD.84.062003},
   DOI={10.1103/physrevd.84.062003},
   number={6},
   journal={Physical Review D},
   publisher={American Physical Society (APS)},
   author={Cornish, Neil and Sampson, Laura and Yunes, Nicolás and Pretorius, Frans},
   year={2011},
   month=sep }

@article{Pratten_2021,
   title={Computationally efficient models for the dominant and subdominant harmonic modes of precessing binary black holes},
   volume={103},
   ISSN={2470-0029},
   url={http://dx.doi.org/10.1103/PhysRevD.103.104056},
   DOI={10.1103/physrevd.103.104056},
   number={10},
   journal={Physical Review D},
   publisher={American Physical Society (APS)},
   author={Pratten, Geraint and García-Quirós, Cecilio and Colleoni, Marta and Ramos-Buades, Antoni and Estellés, Héctor and Mateu-Lucena, Maite and Jaume, Rafel and Haney, Maria and Keitel, David and Thompson, Jonathan E. and Husa, Sascha},
   year={2021},
   month=may }

@article{Janquart_2023,
   title={The return of GOLUM: improving distributed joint parameter estimation for strongly lensed gravitational waves},
   volume={526},
   ISSN={1365-2966},
   url={http://dx.doi.org/10.1093/mnras/stad2838},
   DOI={10.1093/mnras/stad2838},
   number={2},
   journal={Monthly Notices of the Royal Astronomical Society},
   publisher={Oxford University Press (OUP)},
   author={Janquart, Justin and Haris, K and Hannuksela, Otto A and Van Den Broeck, Chris},
   year={2023},
   month=sep, pages={3088–3098} }

@ARTICLE{2025MNRAS.536.2212P,
       author = {{Poon}, Jason S.~C. and {Rinaldi}, Stefano and {Janquart}, Justin and {Narola}, Harsh and {Hannuksela}, Otto A.},
        title = "{Galaxy lens reconstruction based on strongly lensed gravitational waves: similarity transformation degeneracy and mass-sheet degeneracy}",
      journal = {\mnras},
         year = 2025,
        month = jan,
       volume = {536},
       number = {3},
        pages = {2212-2233},
          doi = {10.1093/mnras/stae2660},
archivePrefix = {arXiv},
       eprint = {2406.06463},
 primaryClass = {astro-ph.HE},
       adsurl = {https://ui.adsabs.harvard.edu/abs/2025MNRAS.536.2212P}
}

@article{Seo_2024,
doi = {10.3847/1538-4357/ad35bb},
url = {https://dx.doi.org/10.3847/1538-4357/ad35bb},
year = {2024},
month = {apr},
publisher = {The American Astronomical Society},
volume = {966},
number = {1},
pages = {107},
author = {Seo, Eungwang and Li, Tjonnie G. F. and Hendry, Martin A.},
title = {Inferring Properties of Dark Galactic Halos Using Strongly Lensed Gravitational Waves},
journal = {The Astrophysical Journal}
}

@article{PhysRevLett.130.261401,
  title = {Cosmography Using Strongly Lensed Gravitational Waves from Binary Black Holes},
  author = {Jana, Souvik and Kapadia, Shasvath J. and Venumadhav, Tejaswi and Ajith, Parameswaran},
  journal = {Phys. Rev. Lett.},
  volume = {130},
  issue = {26},
  pages = {261401},
  numpages = {8},
  year = {2023},
  month = {Jun},
  publisher = {American Physical Society},
  doi = {10.1103/PhysRevLett.130.261401},
  url = {https://link.aps.org/doi/10.1103/PhysRevLett.130.261401}
}

@ARTICLE{2022A&A...659L...5C,
       author = {{Cao}, Shuo and {Qi}, Jingzhao and {Cao}, Zhoujian and {Biesiada}, Marek and {Cheng}, Wei and {Zhu}, Zong-Hong},
        title = "{Direct measurement of the distribution of dark matter with strongly lensed gravitational waves}",
      journal = {\aap},
         year = 2022,
        month = mar,
       volume = {659},
          eid = {L5},
        pages = {L5},
          doi = {10.1051/0004-6361/202142694},
archivePrefix = {arXiv},
       eprint = {2202.08714},
 primaryClass = {astro-ph.CO},
       adsurl = {https://ui.adsabs.harvard.edu/abs/2022A&A...659L...5C}
}

@ARTICLE{2020LRR....23....3A,
       author = {{Abbott}, B.~P. and {Abbott}, R. and {Abbott}, T.~D. and {Abraham}, S. and {Acernese}, F. and {Ackley}, K. and {Adams}, C. and {Adya}, V.~B. and {Affeldt}, C. and {Agathos}, M. and {Agatsuma}, K. and {Aggarwal}, N. and {Aguiar}, O.~D. and {Aiello}, L. and {Ain}, A. and {Ajith}, P. and {Akutsu}, T. and {Allen}, G. and {Allocca}, A. and {Aloy}, M.~A. and {Altin}, P.~A. and {Amato}, A. and {Ananyeva}, A. and {Anderson}, S.~B. and {Anderson}, W.~G. and {Ando}, M. and {Angelova}, S.~V. and {Antier}, S. and {Appert}, S. and {Arai}, K. and {Arai}, Koya and {Arai}, Y. and {Araki}, S. and {Araya}, A. and {Araya}, M.~C. and {Areeda}, J.~S. and {Ar{\`e}ne}, M. and {Aritomi}, N. and {Arnaud}, N. and {Arun}, K.~G. and {Ascenzi}, S. and {Ashton}, G. and {Aso}, Y. and {Aston}, S.~M. and {Astone}, P. and {Aubin}, F. and {Aufmuth}, P. and {Aultoneal}, K. and {Austin}, C. and {Avendano}, V. and {Avila-Alvarez}, A. and {Babak}, S. and {Bacon}, P. and {Badaracco}, F. and {Bader}, M.~K.~M. and {Bae}, S.~W. and {Bae}, Y.~B. and {Baiotti}, L. and {Bajpai}, R. and {Baker}, P.~T. and {Baldaccini}, F. and {Ballardin}, G. and {Ballmer}, S.~W. and {Banagiri}, S. and {Barayoga}, J.~C. and {Barclay}, S.~E. and {Barish}, B.~C. and {Barker}, D. and {Barkett}, K. and {Barnum}, S. and {Barone}, F. and {Barr}, B. and {Barsotti}, L. and {Barsuglia}, M. and {Barta}, D. and {Bartlett}, J. and {Barton}, M.~A. and {Bartos}, I. and {Bassiri}, R. and {Basti}, A. and {Bawaj}, M. and {Bayley}, J.~C. and {Bazzan}, M. and {B{\'e}csy}, B. and {Bejger}, M. and {Belahcene}, I. and {Bell}, A.~S. and {Beniwal}, D. and {Berger}, B.~K. and {Bergmann}, G. and {Bernuzzi}, S. and {Bero}, J.~J. and {Berry}, C.~P.~L. and {Bersanetti}, D. and {Bertolini}, A. and {Betzwieser}, J. and {Bhandare}, R. and {Bidler}, J. and {Bilenko}, I.~A. and {Bilgili}, S.~A. and {Billingsley}, G. and {Birch}, J. and {Birney}, R. and {Birnholtz}, O. and {Biscans}, S. and {Biscoveanu}, S. and {Bisht}, A. and {Bitossi}, M. and {Bizouard}, M.~A. and {Blackburn}, J.~K. and {Blair}, C.~D. and {Blair}, D.~G. and {Blair}, R.~M. and {Bloemen}, S. and {Bode}, N. and {Boer}, M. and {Boetzel}, Y. and {Bogaert}, G. and {Bondu}, F. and {Bonilla}, E. and {Bonnand}, R. and {Booker}, P. and {Boom}, B.~A. and {Booth}, C.~D. and {Bork}, R. and {Boschi}, V. and {Bose}, S. and {Bossie}, K. and {Bossilkov}, V. and {Bosveld}, J. and {Bouffanais}, Y. and {Bozzi}, A. and {Bradaschia}, C. and {Brady}, P.~R. and {Bramley}, A. and {Branchesi}, M. and {Brau}, J.~E. and {Briant}, T. and {Briggs}, J.~H. and {Brighenti}, F. and {Brillet}, A. and {Brinkmann}, M. and {Brisson}, V. and {Brockill}, P. and {Brooks}, A.~F. and {Brown}, D.~A. and {Brown}, D.~D. and {Brunett}, S. and {Buikema}, A. and {Bulik}, T. and {Bulten}, H.~J. and {Buonanno}, A. and {Buskulic}, D. and {Buy}, C. and {Byer}, R.~L. and {Cabero}, M. and {Cadonati}, L. and {Cagnoli}, G. and {Cahillane}, C. and {Bustillo}, J. Calder{\'o}n and {Callister}, T.~A. and {Calloni}, E. and {Camp}, J.~B. and {Campbell}, W.~A. and {Canepa}, M. and {Cannon}, K. and {Cannon}, K.~C. and {Cao}, H. and {Cao}, J. and {Capocasa}, E. and {Carbognani}, F. and {Caride}, S. and {Carney}, M.~F. and {Carullo}, G. and {Diaz}, J. Casanueva and {Casentini}, C. and {Caudill}, S. and {Cavagli{\`a}}, M. and {Cavalier}, F. and {Cavalieri}, R. and {Cella}, G. and {Cerd{\'a}-Dur{\'a}n}, P. and {Cerretani}, G. and {Cesarini}, E. and {Chaibi}, O. and {Chakravarti}, K. and {Chamberlin}, S.~J. and {Chan}, M. and {Chan}, M.~L. and {Chao}, S. and {Charlton}, P. and {Chase}, E.~A. and {Chassande-Mottin}, E. and {Chatterjee}, D. and {Chaturvedi}, M. and {Chatziioannou}, K. and {Cheeseboro}, B.~D. and {Chen}, C.~S. and {Chen}, H.~Y. and {Chen}, K.~H.},
        title = "{Prospects for observing and localizing gravitational-wave transients with Advanced LIGO, Advanced Virgo and KAGRA}",
      journal = {Living Reviews in Relativity},
         year = 2020,
        month = dec,
       volume = {23},
       number = {1},
          eid = {3},
        pages = {3},
          doi = {10.1007/s41114-020-00026-9},
       adsurl = {https://ui.adsabs.harvard.edu/abs/2020LRR....23....3A}
}

@article{Hu:2025vlp,
    author = "Hu, Qian",
    title = "{Hierarchical Subtraction with Neural Density Estimators as a General Solution to Overlapping Gravitational Wave Signals}",
    eprint = "2507.05209",
    archivePrefix = "arXiv",
    primaryClass = "gr-qc",
    reportNumber = "LIGO-P2500409, ET-0346A-25",
    month = "7",
    year = "2025"
}

@article{Seo:2025dto,
    author = "Seo, Eungwang and Shan, Xikai and Janquart, Justin and Hannuksela, Otto A. and Hendry, Martin A. and Hu, Bin",
    title = "{Residual Test to Search for Microlensing Signatures in Strongly Lensed Gravitational Wave Signals}",
    eprint = "2503.02186",
    archivePrefix = "arXiv",
    primaryClass = "gr-qc",
    doi = "10.3847/1538-4357/ade4bf",
    journal = "Astrophys. J.",
    volume = "988",
    number = "2",
    pages = "159",
    year = "2025"
}

@article{Nakamura:1997sw,
    author = "Nakamura, Takahiro T.",
    title = "{Gravitational lensing of gravitational waves from inspiraling binaries by a point mass lens}",
    reportNumber = "UTAP-272-97, YITP-97-61",
    doi = "10.1103/PhysRevLett.80.1138",
    journal = "Phys. Rev. Lett.",
    volume = "80",
    pages = "1138--1141",
    year = "1998"
}

@article{Takahashi:2003ix,
    author = "Takahashi, Ryuichi and Nakamura, Takashi",
    title = "{Wave effects in gravitational lensing of gravitational waves from chirping binaries}",
    eprint = "astro-ph/0305055",
    archivePrefix = "arXiv",
    doi = "10.1086/377430",
    journal = "Astrophys. J.",
    volume = "595",
    pages = "1039--1051",
    year = "2003"
}

@article{Diego:2019lcd,
    author = "Diego, J. M. and Hannuksela, O. A. and Kelly, P. L. and Broadhurst, T. and Kim, K. and Li, T. G. F. and Smoot, G. F. and Pagano, G.",
    title = "{Observational signatures of microlensing in gravitational waves at LIGO/Virgo frequencies}",
    eprint = "1903.04513",
    archivePrefix = "arXiv",
    primaryClass = "astro-ph.CO",
    doi = "10.1051/0004-6361/201935490",
    journal = "Astron. Astrophys.",
    volume = "627",
    pages = "A130",
    year = "2019"
}

@article{Mishra:2021xzz,
    author = "Mishra, Anuj and Meena, Ashish Kumar and More, Anupreeta and Bose, Sukanta and Bagla, Jasjeet Singh",
    title = "{Gravitational lensing of gravitational waves: effect of microlens population in lensing galaxies}",
    eprint = "2102.03946",
    archivePrefix = "arXiv",
    primaryClass = "astro-ph.CO",
    doi = "10.1093/mnras/stab2875",
    journal = "Mon. Not. Roy. Astron. Soc.",
    volume = "508",
    number = "4",
    pages = "4869--4886",
    year = "2021"
}

@article{Liu:2023emk,
    author = "Liu, Anna and Kim, Kyungmin",
    title = "{Can we discern millilensed gravitational-wave signals from signals produced by precessing binary black holes with ground-based detectors?}",
    eprint = "2301.07253",
    archivePrefix = "arXiv",
    primaryClass = "gr-qc",
    reportNumber = "LIGO-P2200398",
    doi = "10.1103/PhysRevD.110.123008",
    journal = "Phys. Rev. D",
    volume = "110",
    number = "12",
    pages = "123008",
    year = "2024"
}

@article{LIGOScientific:2016aoc,
    author = "Abbott, B. P. and others",
    collaboration = "LIGO Scientific, Virgo",
    title = "{Observation of Gravitational Waves from a Binary Black Hole Merger}",
    eprint = "1602.03837",
    archivePrefix = "arXiv",
    primaryClass = "gr-qc",
    reportNumber = "LIGO-P150914",
    doi = "10.1103/PhysRevLett.116.061102",
    journal = "Phys. Rev. Lett.",
    volume = "116",
    number = "6",
    pages = "061102",
    year = "2016"
}

@article{Belczynski_2016,
   title={The first gravitational-wave source from the isolated evolution of two stars in the 40–100 solar mass range},
   volume={534},
   ISSN={1476-4687},
   url={http://dx.doi.org/10.1038/nature18322},
   DOI={10.1038/nature18322},
   number={7608},
   journal={Nature},
   publisher={Springer Science and Business Media LLC},
   author={Belczynski, Krzysztof and Holz, Daniel E. and Bulik, Tomasz and O’Shaughnessy, Richard},
   year={2016},
   month=jun, pages={512–515} }

@ARTICLE{2018MNRAS.480.2011G,
       author = {{Giacobbo}, Nicola and {Mapelli}, Michela},
        title = "{The progenitors of compact-object binaries: impact of metallicity, common envelope and natal kicks}",
      journal = {\mnras},
         year = 2018,
        month = oct,
       volume = {480},
       number = {2},
        pages = {2011-2030},
          doi = {10.1093/mnras/sty1999},
archivePrefix = {arXiv},
       eprint = {1806.00001},
 primaryClass = {astro-ph.HE},
       adsurl = {https://ui.adsabs.harvard.edu/abs/2018MNRAS.480.2011G}
}

@ARTICLE{1993Natur.364..423S,
       author = {{Sigurdsson}, Steinn and {Hernquist}, Lars},
        title = "{Primordial black holes in globular clusters}",
      journal = {\nat},
         year = 1993,
        month = jul,
       volume = {364},
       number = {6436},
        pages = {423-425},
          doi = {10.1038/364423a0},
       adsurl = {https://ui.adsabs.harvard.edu/abs/1993Natur.364..423S}
}

@ARTICLE{2000ApJ...528L..17P,
       author = {{Portegies Zwart}, Simon F. and {McMillan}, Stephen L.~W.},
        title = "{Black Hole Mergers in the Universe}",
      journal = {\apjl},
         year = 2000,
        month = jan,
       volume = {528},
       number = {1},
        pages = {L17-L20},
          doi = {10.1086/312422},
archivePrefix = {arXiv},
       eprint = {astro-ph/9910061},
 primaryClass = {astro-ph},
       adsurl = {https://ui.adsabs.harvard.edu/abs/2000ApJ...528L..17P}
}

@article{Rodriguez_2016,
   title={DYNAMICAL FORMATION OF THE GW150914 BINARY BLACK HOLE},
   volume={824},
   ISSN={2041-8213},
   url={http://dx.doi.org/10.3847/2041-8205/824/1/L8},
   DOI={10.3847/2041-8205/824/1/l8},
   number={1},
   journal={The Astrophysical Journal Letters},
   publisher={American Astronomical Society},
   author={Rodriguez, Carl L. and Haster, Carl-Johan and Chatterjee, Sourav and Kalogera, Vicky and Rasio, Frederic A.},
   year={2016},
   month=jun, pages={L8} }

@article{Mapelli_2022,
   title={The cosmic evolution of binary black holes in young, globular, and nuclear star clusters: rates, masses, spins, and mixing fractions},
   volume={511},
   ISSN={1365-2966},
   url={http://dx.doi.org/10.1093/mnras/stac422},
   DOI={10.1093/mnras/stac422},
   number={4},
   journal={Monthly Notices of the Royal Astronomical Society},
   publisher={Oxford University Press (OUP)},
   author={Mapelli, Michela and Bouffanais, Yann and Santoliquido, Filippo and Arca Sedda, Manuel and Artale, M Celeste},
   year={2022},
   month=feb, pages={5797–5816} }

@article{Lo:2021nae,
    author = "Lo, Rico K. L. and Maga\~na Hernandez, Ignacio",
    title = "{A Bayesian statistical framework for identifying strongly-lensed gravitational-wave signals}",
    eprint = "2104.09339",
    archivePrefix = "arXiv",
    primaryClass = "gr-qc",
    month = "4",
    year = "2021"
}

@article{Liu:2020par,
    author = "Liu, Xiaoshu and Magana Hernandez, Ignacio and Creighton, Jolien",
    title = "{Identifying strong gravitational-wave lensing during the second observing run of Advanced LIGO and Advanced Virgo}",
    eprint = "2009.06539",
    archivePrefix = "arXiv",
    primaryClass = "astro-ph.HE",
    doi = "10.3847/1538-4357/abd7eb",
    journal = "Astrophys. J.",
    volume = "908",
    number = "1",
    pages = "97",
    year = "2021"
}

@article{Choi_2007,
   title={Internal and Collective Properties of Galaxies in the Sloan Digital Sky Survey},
   volume={658},
   ISSN={1538-4357},
   url={http://dx.doi.org/10.1086/511060},
   DOI={10.1086/511060},
   number={2},
   journal={The Astrophysical Journal},
   publisher={American Astronomical Society},
   author={Choi, Yun‐Young and Park, Changbom and Vogeley, Michael S.},
   year={2007},
   month=apr, pages={884–897} }

@article{PhysRevD.101.064011,
  title = {Gravitational wave interference via gravitational lensing: Measurements of luminosity distance, lens mass, and cosmological parameters},
  author = {Hou, Shaoqi and Fan, Xi-Long and Liao, Kai and Zhu, Zong-Hong},
  journal = {Phys. Rev. D},
  volume = {101},
  issue = {6},
  pages = {064011},
  numpages = {9},
  year = {2020},
  month = {Mar},
  publisher = {American Physical Society},
  doi = {10.1103/PhysRevD.101.064011},
  url = {https://link.aps.org/doi/10.1103/PhysRevD.101.064011}
}

@article{Lai:2018rto,
    author = "Lai, Kwun-Hang and Hannuksela, Otto A. and Herrera-Mart\'\i{}n, Antonio and Diego, Jose M. and Broadhurst, Tom and Li, Tjonnie G. F.",
    title = "{Discovering intermediate-mass black hole lenses through gravitational wave lensing}",
    eprint = "1801.07840",
    archivePrefix = "arXiv",
    primaryClass = "gr-qc",
    doi = "10.1103/PhysRevD.98.083005",
    journal = "Phys. Rev. D",
    volume = "98",
    number = "8",
    pages = "083005",
    year = "2018"
}

@BOOK{1992grlebookS,
       author = {{Schneider}, Peter and {Ehlers}, J{\"u}rgen and {Falco}, Emilio E.},
        title = "{Gravitational Lenses}",
         year = 1992,
          doi = {10.1007/978-3-662-03758-4},
       adsurl = {https://ui.adsabs.harvard.edu/abs/1992grle.book.....S}
}

@article{2021xuechunchen,
   title={FRBs Lensed by Point Masses. II. The Multipeaked FRBs from the Point View of Microlensing},
   volume={923},
   ISSN={1538-4357},
   url={http://dx.doi.org/10.3847/1538-4357/ac2c76},
   DOI={10.3847/1538-4357/ac2c76},
   number={1},
   journal={The Astrophysical Journal},
   publisher={American Astronomical Society},
   author={Chen, Xuechun and Shu, Yiping and Li, Guoliang and Zheng, Wenwen},
   year={2021},
   month={Dec},
   pages={117} 
}

@incollection{schneider1992gravitational,
  title={Gravitational lenses as astrophysical tools},
  author={Schneider, Peter and Ehlers, J{\"u}rgen and Falco, Emilio E},
  booktitle={Gravitational Lenses},
  pages={467--515},
  year={1992},
  publisher={Springer}
}

@article{10.1143/PTPS.133.137,
    author = {Nakamura, Takahiro T. and Deguchi, Shuji},
    title = "{Wave Optics in Gravitational Lensing}",
    journal = {Progress of Theoretical Physics Supplement},
    volume = {133},
    pages = {137-153},
    year = {1999},
    month = {01},
    issn = {0375-9687},
    doi = {10.1143/PTPS.133.137},
    url = {https://doi.org/10.1143/PTPS.133.137},
    eprint = {https://academic.oup.com/ptps/article-pdf/doi/10.1143/PTPS.133.137/5283012/133-137.pdf},
}

@PHDTHESIS{Wambsganss1990,
       author = {{Wambsganss}, J.},
        title = {Gravitational Microlensing},
        school = {Munich University},
         year = {1990},
        type  = {Dissertation for Doctoral Degree},
       adsurl = {https://ui.adsabs.harvard.edu/abs/1990PhDT.......180W}
}

@article{Janquart:2021nus,
    author = "Janquart, Justin and Seo, Eungwang and Hannuksela, Otto A. and Li, Tjonnie G. F. and Broeck, Chris Van Den",
    title = "{On the Identification of Individual Gravitational-wave Image Types of a Lensed System Using Higher-order Modes}",
    eprint = "2110.06873",
    archivePrefix = "arXiv",
    primaryClass = "gr-qc",
    doi = "10.3847/2041-8213/ac3bcf",
    journal = "Astrophys. J. Lett.",
    volume = "923",
    number = "1",
    pages = "L1",
    year = "2021"
}

@misc{zheng2022,
  doi = {10.48550/ARXIV.2204.10871},
  
  url = {https://arxiv.org/abs/2204.10871},
  
  author = {Zheng, Wenwen and Chen, Xuechun and Li, Guoliang and Chen, Hou-Zun},
  
  title = {An Improved GPU-Based Ray-Shooting Code For Gravitational Microlensing},
  
  publisher = {arXiv},
  
  year = {2022},
  
  copyright = {Creative Commons Attribution Non Commercial Share Alike 4.0 International}
}

@article{Shan:2022xfx,
    author = "Shan, Xikai and Li, Guoliang and Chen, Xuechun and Zheng, Wenwen and Zhao, Wen",
    title = "{Wave effect of gravitational waves intersected with a microlens field: A new algorithm and supplementary study}",
    eprint = "2208.13566",
    archivePrefix = "arXiv",
    primaryClass = "astro-ph.CO",
    doi = "10.1007/s11433-022-1985-3",
    journal = "Sci. China Phys. Mech. Astron.",
    volume = "66",
    number = "3",
    pages = "239511",
    year = "2023"
}

@article{Oguri:2019fix,
    author = "Oguri, Masamune",
    title = "{Strong gravitational lensing of explosive transients}",
    eprint = "1907.06830",
    archivePrefix = "arXiv",
    primaryClass = "astro-ph.CO",
    doi = "10.1088/1361-6633/ab4fc5",
    journal = "Rept. Prog. Phys.",
    volume = "82",
    number = "12",
    pages = "126901",
    year = "2019"
}

@article{Liao:2022gde,
    author = "Liao, Kai and Biesiada, Marek and Zhu, Zong-Hong",
    title = "{Strongly Lensed Transient Sources: A Review}",
    eprint = "2207.13489",
    archivePrefix = "arXiv",
    primaryClass = "astro-ph.HE",
    doi = "10.1088/0256-307X/39/11/119801",
    journal = "Chin. Phys. Lett.",
    volume = "39",
    number = "11",
    pages = "119801",
    year = "2022"
}

@article{Caliskan:2023zqm,
    author = "\c{C}al\i{}\c{s}kan, Mesut and Anil Kumar, Neha and Ji, Lingyuan and Ezquiaga, Jose M. and Cotesta, Roberto and Berti, Emanuele and Kamionkowski, Marc",
    title = "{Probing wave-optics effects and low-mass dark matter halos with lensing of gravitational waves from massive black holes}",
    eprint = "2307.06990",
    archivePrefix = "arXiv",
    primaryClass = "astro-ph.CO",
    doi = "10.1103/PhysRevD.108.123543",
    journal = "Phys. Rev. D",
    volume = "108",
    number = "12",
    pages = "123543",
    year = "2023"
}

@article{Shan:2023qvd,
    author = "Shan, Xikai and Chen, Xuechun and Hu, Bin and Li, Guoliang",
    title = "{Microlensing bias on the detection of strong lensing gravitational wave}",
    eprint = "2306.14796",
    archivePrefix = "arXiv",
    primaryClass = "astro-ph.CO",
    doi = "10.1007/s11433-023-2334-9",
    journal = "Sci. China Phys. Mech. Astron.",
    volume = "67",
    number = "6",
    pages = "269511",
    year = "2024"
}

@article{Shan:2023ngi,
    author = "Shan, Xikai and Chen, Xuechun and Hu, Bin and Cai, Rong-Gen",
    title = "{Microlensing sheds light on the detection of strong lensing gravitational waves}",
    eprint = "2301.06117",
    archivePrefix = "arXiv",
    primaryClass = "astro-ph.IM",
    month = "1",
    year = "2023",
    url = {https://arxiv.org/abs/2301.06117}
    
}

@article{Haris:2018vmn,
    author = "Haris, K. and Mehta, Ajit Kumar and Kumar, Sumit and Venumadhav, Tejaswi and Ajith, Parameswaran",
    title = "{Identifying strongly lensed gravitational wave signals from binary black hole mergers}",
    eprint = "1807.07062",
    archivePrefix = "arXiv",
    primaryClass = "gr-qc",
    reportNumber = "LIGO- P1800155",
    month = "7",
    year = "2018",
    url = {https://arxiv.org/abs/1807.07062}
}

@article{Xu:2021bfn,
    author = "Xu, Fei and Ezquiaga, Jose Maria and Holz, Daniel E.",
    title = "{Please Repeat: Strong Lensing of Gravitational Waves as a Probe of Compact Binary and Galaxy Populations}",
    eprint = "2105.14390",
    archivePrefix = "arXiv",
    primaryClass = "astro-ph.CO",
    doi = "10.3847/1538-4357/ac58f8",
    journal = "Astrophys. J.",
    volume = "929",
    number = "1",
    pages = "9",
    year = "2022"
}

@ARTICLE{2019S&C....29..891H,
       author = {{Higson}, Edward and {Handley}, Will and {Hobson}, Mike and {Lasenby}, Anthony},
        title = "{Dynamic nested sampling: an improved algorithm for parameter estimation and evidence calculation}",
      journal = {Statistics and Computing},
         year = 2019,
        month = sep,
       volume = {29},
       number = {5},
        pages = {891-913},
          doi = {10.1007/s11222-018-9844-0},
archivePrefix = {arXiv},
       eprint = {1704.03459},
 primaryClass = {stat.CO},
       adsurl = {https://ui.adsabs.harvard.edu/abs/2019S&C....29..891H}
}

@article{Shan:2024min,
    author = "Shan, Xikai and Li, Guoliang and Chen, Xuechun and Zhao, Wen and Hu, Bin and Mao, Shude",
    title = "{Wave effect of gravitational waves intersected with a microlens field II: An adaptive hierarchical tree algorithm and population study}",
    eprint = "2409.06747",
    archivePrefix = "arXiv",
    primaryClass = "astro-ph.IM",
    doi = "10.1007/s11433-024-2502-1",
    journal = "Sci. China Phys. Mech. Astron.",
    volume = "68",
    number = "1",
    pages = "219512",
    year = "2025"
}

@article{LIGOScientific:2018jsj,
    author = "Abbott, B. P. and others",
    collaboration = "LIGO Scientific, Virgo",
    title = "{Binary Black Hole Population Properties Inferred from the First and Second Observing Runs of Advanced LIGO and Advanced Virgo}",
    eprint = "1811.12940",
    archivePrefix = "arXiv",
    primaryClass = "astro-ph.HE",
    reportNumber = "LIGO-P1800324",
    doi = "10.3847/2041-8213/ab3800",
    journal = "Astrophys. J. Lett.",
    volume = "882",
    number = "2",
    pages = "L24",
    year = "2019"
}

@ARTICLE{1994A&A...284..285K,
       author = {{Kormann}, R. and {Schneider}, P. and {Bartelmann}, M.},
        title = "{Isothermal elliptical gravitational lens models.}",
      journal = {\aap},
         year = 1994,
        month = apr,
       volume = {284},
        pages = {285-299},
       adsurl = {https://ui.adsabs.harvard.edu/abs/1994A&A...284..285K}
}

@article{Wierda:2021upe,
    author = "Wierda, A. Renske A. C. and Wempe, Ewoud and Hannuksela, Otto A. and Koopmans, L. \'eon V. E. and Van Den Broeck, Chris",
    title = "{Beyond the Detector Horizon: Forecasting Gravitational-Wave Strong Lensing}",
    eprint = "2106.06303",
    archivePrefix = "arXiv",
    primaryClass = "astro-ph.HE",
    doi = "10.3847/1538-4357/ac1bb4",
    journal = "Astrophys. J.",
    volume = "921",
    number = "2",
    pages = "154",
    year = "2021"
}

@ARTICLE{2003PASP..115..763C,
       author = {{Chabrier}, Gilles},
        title = "{Galactic Stellar and Substellar Initial Mass Function}",
      journal = {\pasp},
         year = 2003,
        month = jul,
       volume = {115},
       number = {809},
        pages = {763-795},
          doi = {10.1086/376392},
archivePrefix = {arXiv},
       eprint = {astro-ph/0304382},
 primaryClass = {astro-ph},
       adsurl = {https://ui.adsabs.harvard.edu/abs/2003PASP..115..763C}
}

@article{Vernardos_2018,
    doi = {10.1093/mnras/sty3486},
  
    url = {https://doi.org/10.1093%2Fmnras%2Fsty3486},
  
    year = 2018,
    month = {dec},
  
    publisher = {Oxford University Press ({OUP})},
  
    volume = {483},
  
    number = {4},
  
    pages = {5583--5594},
  
    author = {G Vernardos},
  
    title = {Microlensing flux ratio predictions for Euclid},
  
    journal = {Monthly Notices of the Royal Astronomical Society}
}

@article{Diego:2021mhf,
    author = {Diego, J. M. and Bernstein, G. and Chen, W. and Goobar, A. and Johansson, J. P. and Kelly, P. L. and M\"ortsell, E. and Nightingale, J. W.},
    title = "{Microlensing and the type Ia supernova iPTF16geu}",
    eprint = "2112.04524",
    archivePrefix = "arXiv",
    primaryClass = "astro-ph.CO",
    doi = "10.1051/0004-6361/202143009",
    journal = "Astron. Astrophys.",
    volume = "662",
    pages = "A34",
    year = "2022"
}

@ARTICLE{2015MNRAS.451.4086S,
       author = {{Spera}, Mario and {Mapelli}, Michela and {Bressan}, Alessandro},
        title = "{The mass spectrum of compact remnants from the PARSEC stellar evolution tracks}",
      journal = {\mnras},
         year = 2015,
        month = aug,
       volume = {451},
       number = {4},
        pages = {4086-4103},
          doi = {10.1093/mnras/stv1161},
archivePrefix = {arXiv},
       eprint = {1505.05201},
 primaryClass = {astro-ph.SR},
       adsurl = {https://ui.adsabs.harvard.edu/abs/2015MNRAS.451.4086S}
}

@article{Meena:2022unp,
    author = "Meena, Ashish Kumar and Mishra, Anuj and More, Anupreeta and Bose, Sukanta and Bagla, Jasjeet Singh",
    title = "{Gravitational lensing of gravitational waves: Probability of microlensing in galaxy-scale lens population}",
    eprint = "2205.05409",
    archivePrefix = "arXiv",
    primaryClass = "astro-ph.GA",
    doi = "10.1093/mnras/stac2721",
    journal = "Mon. Not. Roy. Astron. Soc.",
    volume = "517",
    number = "1",
    pages = "872--884",
    year = "2022"
}

@article{Li:2018prc,
    author = "Li, Shun-Sheng and Mao, Shude and Zhao, Yuetong and Lu, Youjun",
    title = "{Gravitational lensing of gravitational waves: A statistical perspective}",
    eprint = "1802.05089",
    archivePrefix = "arXiv",
    primaryClass = "astro-ph.CO",
    doi = "10.1093/mnras/sty411",
    journal = "Mon. Not. Roy. Astron. Soc.",
    volume = "476",
    number = "2",
    pages = "2220--2229",
    year = "2018"
}

@article{Hannuksela:2020xor,
    author = "Hannuksela, Otto A. and Collett, Thomas E. and \c{C}al\i{}\c{s}kan, Mesut and Li, Tjonnie G. F.",
    title = "{Localizing merging black holes with sub-arcsecond precision using gravitational-wave lensing}",
    eprint = "2004.13811",
    archivePrefix = "arXiv",
    primaryClass = "astro-ph.HE",
    doi = "10.1093/mnras/staa2577",
    journal = "Mon. Not. Roy. Astron. Soc.",
    volume = "498",
    number = "3",
    pages = "3395--3402",
    year = "2020"
}

@article{Liu:2023ikc,
    author = "Liu, Anna and Wong, Isaac C. F. and Leong, Samson H. W. and More, Anupreeta and Hannuksela, Otto A. and Li, Tjonnie G. F.",
    title = "{Exploring the hidden Universe: a novel phenomenological approach for recovering arbitrary gravitational-wave millilensing configurations}",
    eprint = "2302.09870",
    archivePrefix = "arXiv",
    primaryClass = "gr-qc",
    doi = "10.1093/mnras/stad1302",
    journal = "Mon. Not. Roy. Astron. Soc.",
    volume = "525",
    number = "3",
    pages = "4149--4160",
    year = "2023"
}

@ARTICLE{1986ApJ...306....2K,
       author = {{Katz}, N. and {Balbus}, S. and {Paczynski}, B.},
        title = "{Random Scattering Approach to Gravitational Microlensing}",
      journal = {\apj},
         year = 1986,
        month = jul,
       volume = {306},
        pages = {2},
          doi = {10.1086/164313},
       adsurl = {https://ui.adsabs.harvard.edu/abs/1986ApJ...306....2K}
}
\bibliographystyle{aasjournal}



\end{document}